\documentclass[12pt]{article}  
\usepackage{a4,epsfig}
\def\pd{\phantom{-}}

\def\ifmath#1{\relax\ifmmode #1\else $#1$\fi}%

\def\TeV{\ifmmode {\,\mathrm{ Te\kern -0.1em V}}\else
                   \textrm{Te\kern -0.1em V}\fi}%
\def\GeV{\ifmmode {\,\mathrm{ Ge\kern -0.1em V}}\else
                   \textrm{Ge\kern -0.1em V}\fi}%
\def\MeV{\ifmmode {\,\mathrm{ Me\kern -0.1em V}}\else
                   \textrm{Me\kern -0.1em V}\fi}%
\def\keV{\ifmmode {\,\mathrm{ ke\kern -0.1em V}}\else
                   \textrm{ke\kern -0.1em V}\fi}%
\def\eV{\ifmmode  {\,\mathrm{ e\kern -0.1em V}}\else
                   \textrm{e\kern -0.1em V}\fi}%

\newcommand{\pbi}{\,\rm{pb}^{-1}}

\newcommand{\ff}    {{\rm f}\overline{\rm f}}

\newcommand{\ee}    {\mathrm{e}^+\mathrm{e}^-}

\newcommand{\pp}    {\mathrm{p}\bar{\mathrm{p}}}

\newcommand{\ALR}    {A_{\mathrm{LR}}}

\newcommand{\MH}      {m_{\mathrm{H}}}
\newcommand{\MZ}      {m_{\mathrm{Z}}}
\newcommand{\MW}      {m_{\mathrm{W}}}

\newcommand{\GW}      {\Gamma_{\mathrm{W}}}
\newcommand{\BrWl}    {\mathrm{Br(W \rightarrow \ell\nu)}}

\newcommand{\Ghad}       {\Gamma_{\mathrm{had}}}

\newcommand {\so}   {\sigma_0^{\rm{had}}}

\newcommand {\cAb} {\mbox{$\cal A_{\rm b}$}}
\newcommand {\cAc} {\mbox{$\cal A_{\rm c}$}}

\newcommand{\swsqeffl}    {\sin^2\!\theta_{\rm{eff}}^\ell}

\begin{document}



\begin{titlepage}

\pagenumbering{arabic}
\vspace*{-1.5cm}
\begin{tabular*}{15.cm}{lc@{\extracolsep{\fill}}r}
{\bf DELPHI Collaboration} & 
\hspace*{1.3cm} 
\epsfig{figure=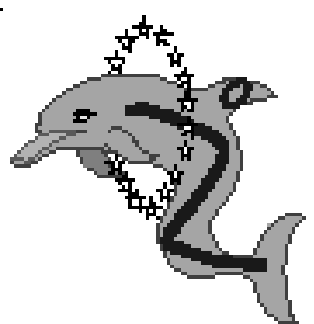,width=1.2cm,height=1.2cm}
&
DELPHI 99-173 TALK 10
\\
& &
22 November 1999
\\
&&\\ \hline
\end{tabular*}
\vspace*{2.cm}
\begin{center}
\Large 
{\bf \boldmath
Electroweak Physics, Experimental Aspects
} \\
\vspace*{2.cm}
\normalsize { 
   {\bf Klaus M\"onig}\\
   {\footnotesize DESY-Zeuthen}\\
   
}
\end{center}
\vspace{\fill}
\begin{abstract}
\noindent
Collider measurements on electroweak physics are summarised. Although the
precision on some observables is very high, no deviation from the Standard
Model of electroweak interactions is observed. The data allow to set stringent
limits on some models for new physics.

\end{abstract}
\vspace{\fill}
\begin{center}
Plenary Talk at the UK Phenomenology Workshop on Collider Physics, Durham, 1999
\end{center}
\vspace{\fill}
\end{titlepage}





\section{Introduction}
Within the last ten years precision tests of the electroweak theory has been
a field of rapid progress. At colliders our knowledge mainly comes
from three sources: $\ee$-interactions close to the peak of the Z-resonance,
$\ee$-interactions above the W-pair production threshold and
$\pp$-interactions above $1 \TeV$ centre of mass energy.
In $\ee$ close to the Z-peak each of the four LEP experiments has
recorded around 4.5 million Z-decays between 1989 and 1995 and their
results are final or nearly final. SLD at SLAC has collected around
half a million events with polarised electron beams until end of 1998
and first results with the full dataset have been presented.
The LEP2 program at energies above $160 \GeV$ is still ongoing. The
experiments have mainly analysed their datasets of around $250 \pbi$
integrated luminosity each at energies up to $189 \GeV$ taken until the end
of 1998. The hope is to more than double the luminosity, mainly at energies
between $192$ and $206 \GeV$.
At the TEVATRON the two experiments have collected each $100 \pbi$ at
$\sqrt{s}=1.8 \TeV$ until 1995. The analysis of this dataset is
basically complete. End of next year the TEVATRON will restart for run II to
collect a factor 20 or more events.

This report summarises the experimental status of electroweak interactions
in the summer 1999. The results in Z-physics, four-fermion physics and
two-fermion physics at high energies are presented and the data are
interpreted within the framework of the Standard Model.

\section{Precision tests on the Z}
Most results from LEP represent only marginal updates with respect to
last year \cite{ref:ew98}. The most interesting change occurred in the
value of the hadronic peak cross section \cite{ref:lepone}. 
The experimental data
remained constant, however due to additional ${\cal O}(\alpha^3)$
corrections for initial state radiation and a new version of the
fitting program ZFITTER \cite{ref:zfitter} the value for 
$\so \left(= \frac{12 \pi}{m_Z} \frac{\Gamma_e \Ghad}{\Gamma_Z^2} \right)$
went up by one standard deviation and due to a decrease in the
theoretical error of the luminosity from $0.1\%$ to $0.06\%$ its error
went down by $30 \%$ so that the that the fitted number of light
neutrino species is now two standard deviations below three
$(N_\nu = 2.9835 \pm 0.0083)$.

SLD have updated their $\ALR$ measurement with the full statistics. 
The result remains stable, however the error decreases by 
$10 \%$ \cite{ref:alr}.
In addition SLD presented several updates of the
left-right-forward-backward asymmetry for b- and c-quarks measuring the final
state coupling parameters $\cAb,\,\cAc$ \cite{ref:abac}. The results are partly
already using the full dataset and the error decreased by $30 \%$ with
respect to last year.

\begin{figure}[htb]
\begin{center}
\includegraphics[height=9.cm,bb=0 88 567 840]{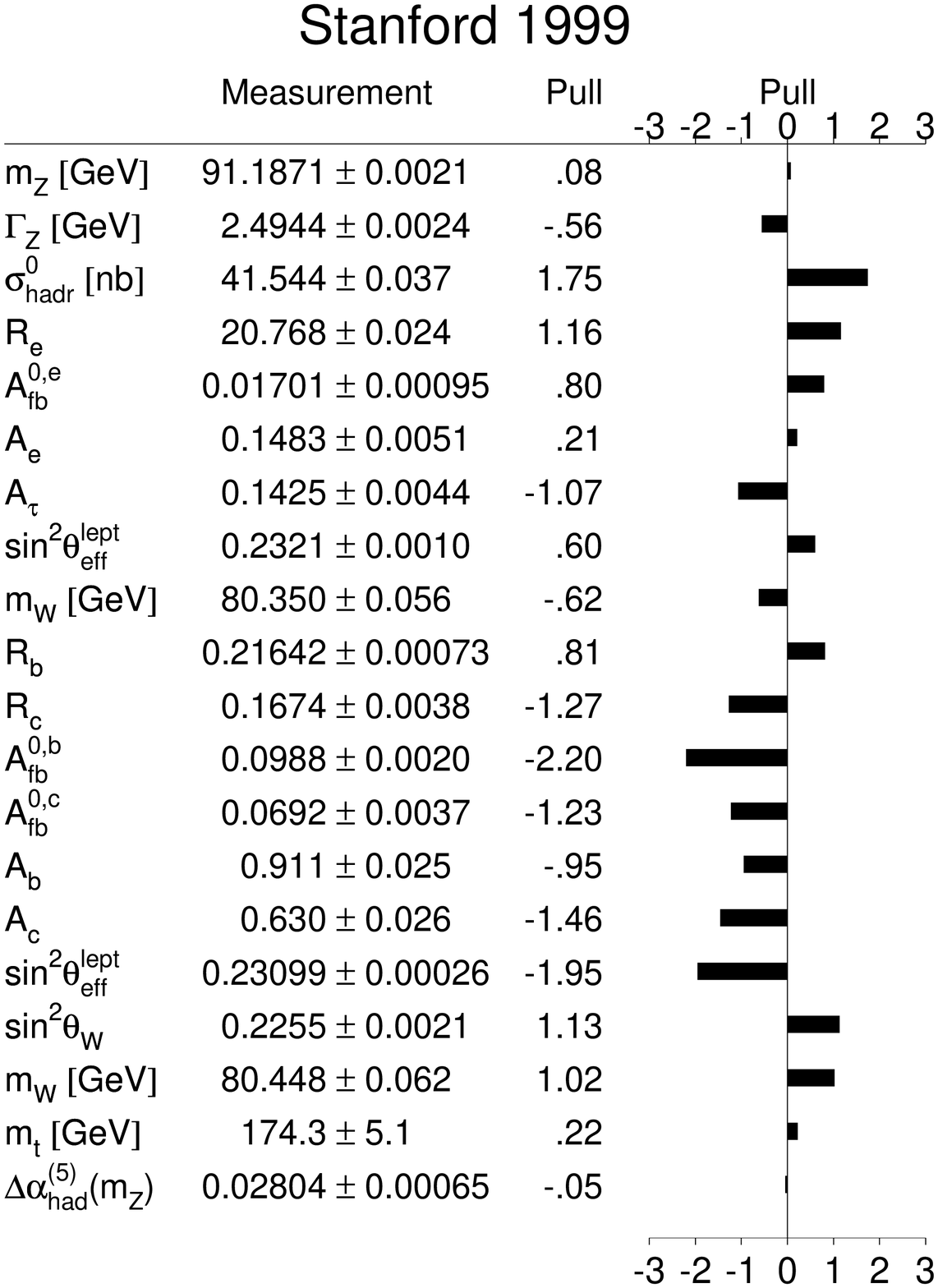}
\includegraphics[height=8.cm,bb=15 35 567 565]{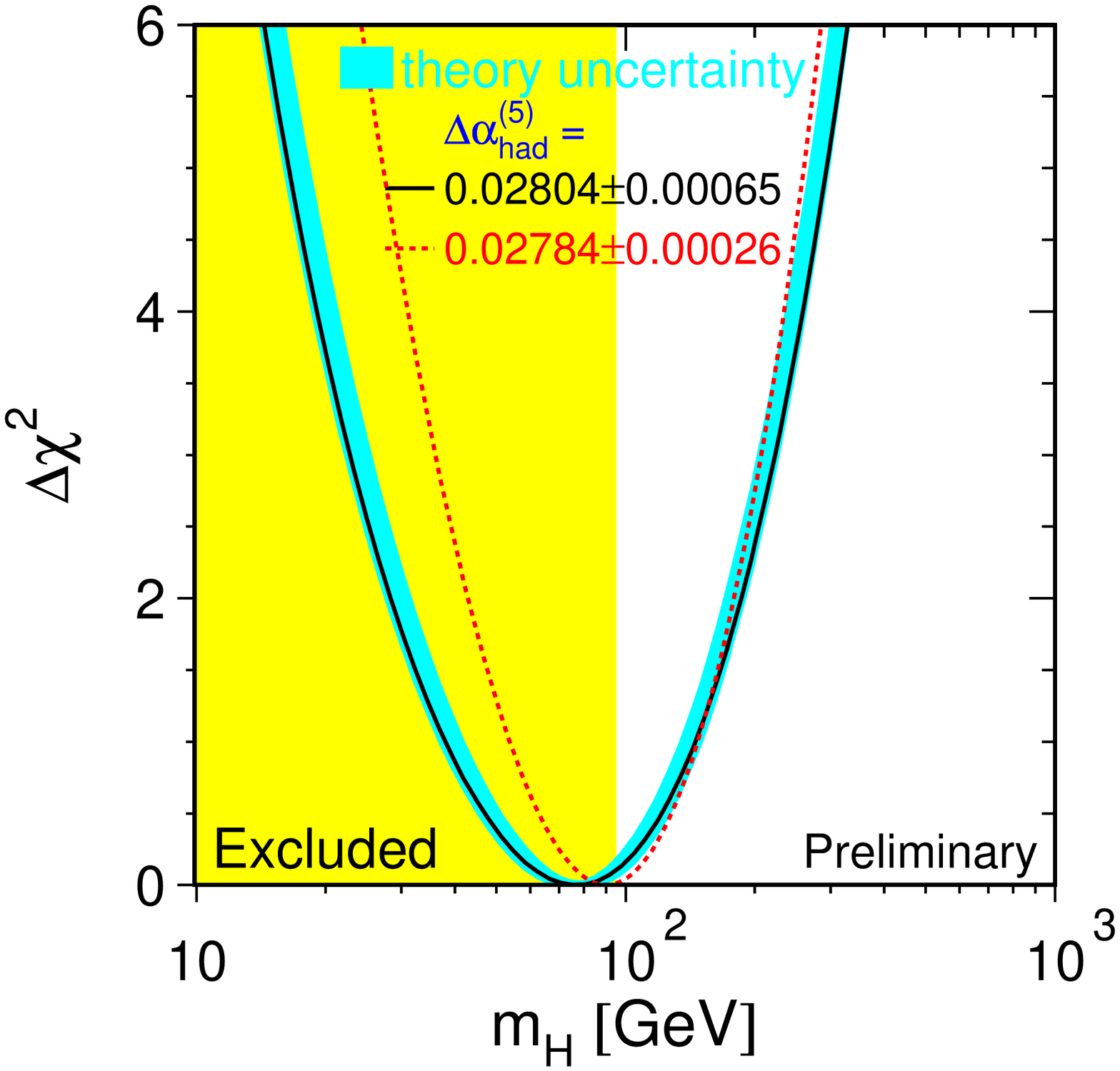}
\end{center}
\caption{a) comparison of the experimental results with the Standard
  Model prediction after the electroweak fit.
  b) $\Delta \chi^2$ of the fit as a function of the Higgs-mass
}
\label{fig:pull}
\end{figure}

Figure \ref{fig:pull} a) summarises all electroweak results and compares
them with the results of the electroweak fit \cite{ref:lepone}. In general good
agreement of the measurements with the Standard Model prediction is
found which is reflected also in $\chi^2/\rm{ndf} = 22.9/15$ corresponding
to a fit probability of $8.6 \%$. 
The two most discrepant observables are the measurement of $\swsqeffl$ from
$\ALR$ at SLD and the b-forward-backward-asymmetry at LEP which both disagree 
by roughly two standard deviations from the fit. These two observables
are responsible for the $2.7 \sigma$ deviation of $\cAb$ obtained
from all relevant LEP and SLD data. It should however be noted that
this is about the worst deviation one can construct from the data and
that the input variables with the largest weight are statistically dominated.

Figure \ref{fig:pull} b) shows the difference in $\chi^2$ of the Standard
Model fit as a function of the Higgs mass for two different values of
$\alpha(\MZ^2)$ \cite{ref:zoltan}. Both fits prefer a Higgs mass in the
$100 \GeV$ region with a $95 \%$ c.l. upper limit of around $220 \GeV$.
The fitted value of $\MH$ is thus consistent with supersymmetric
theories as well as with the Standard Model being valid up to the
Planck mass.

\section{W-physics}
The study of W-properties is one of the major issues at LEP2 and at
the TEVATRON.
At both machines the experiments measure the W-mass and width, its couplings
and branching ratios. At LEP the main source of Ws is W-pair production while 
at the TEVATRON Ws are mainly produced singly in quark antiquark annihilation.

\subsection{The W-mass and branching ratios}
At LEP the W-mass is determined reconstructing the invariant mass of
the final state fermion pairs. For this method the W-pairs with both Ws 
decaying hadronically \cite{ref:mwh} and the ones with one W decaying 
hadronically and the other into a charged lepton and a neutrino 
\cite{ref:mwl} can be used. For
the latter the neutrino momentum is identified with the missing
momentum in the event.
In both cases the resolution is improved by a constrained fit
imposing energy and momentum conservation. All experiments also impose
directly or indirectly the constraint that the masses of the two
pairs are (about) equal.
The constrained fits improve the experimental resolution significantly and
make the mass, reconstructed from the fully hadronic decays,
practically independent of the energy scale of the experiment. On the
other hand they make the results dependent on the knowledge of the beam
energy and on initial state radiation. 
For the fully hadronic decays the analysis is complicated by final state 
effects \cite{ref:fsi}. 
The W flight distance is about 0.1 fm which a a factor 10 smaller
than the typical fragmentation radius. The quarks and gluons from the 
different Ws can therefor interact with each other during fragmentation.
Recent studies on this effect give mass shifts of the order of $50 \MeV$ which
is taken as a systematic error. In addition Bose Einstein correlations between
the final state pions have to be taken into account. It is theoretically
unclear if these correlations should be visible for particles from different
Ws and the experimental situation is contradictory. From model studies an error
of about $20 \MeV$ is taken as systematic error.

At $\pp$ colliders the W mass can be measured only for leptonic W decays
\cite{ref:mwt}. 
Since an unknown longitudinal momentum is lost in the beampipe only the
transverse momentum of the neutrino can be reconstructed. The W-mass is thus 
reconstructed from the transverse mass distribution of the charged 
lepton/neutrino pair. The main limitations at present are the understanding
of the energy scale of the detector and of the hadronic recoil. Both are 
calibrated with leptonic Z decays, so that the corresponding errors are mainly
of statistical nature. The remaining theoretical errors are small at present,
but could become relevant again at run II.

\begin{figure}[htb]
\begin{center}
\includegraphics[height=7.cm,bb=0 105 500 470]{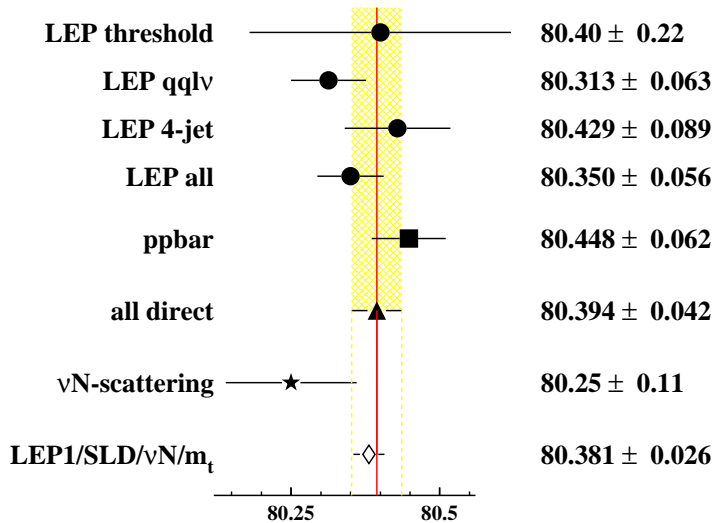}
\end{center}
\caption{
Measurements of $\MW$ using the different methods.
}
\label{fig:mw}
\end{figure}

Figure \ref{fig:mw} summarises the W-masses obtained with the different
methods \cite{ref:mwl,ref:mwh,ref:mwt}. 
It can be seen that the results agree very well and that 
the precision of the direct measurement starts to match the one 
of the prediction from the accurate Z data.

From the W-mass spectra at LEP or from the high end of the transverse
mass spectrum at the TEVATRON also the W width can be
determined. However to date only CDF has done the measurement at the
TEVATRON and only L3 has included the $189 \GeV$ data. Both measurements
(LEP: $\GW = 2.12 \pm 0.20 \GeV$ \cite{ref:gwl}, 
CDF: $\GW = 2.055 \pm 0.125 \GeV$ \cite{ref:mwt})
are in agreement with the SM prediction of $2.08 \GeV$.

\begin{figure}[htb]
\begin{center}
\includegraphics[height=7.cm,bb=55 105 530 480]{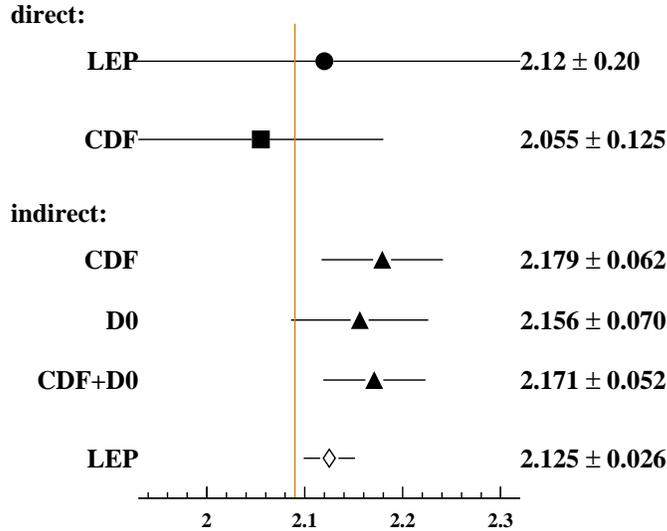}
\end{center}
\caption{
Direct and indirect measurements of $\GW$.
}
\label{fig:gw}
\end{figure}

At LEP the W branching ratios are measured together with the W-pair
cross section. Under the assumption that all W decays are visible, each
W pair event can be classified according to the two decays and the
branching ratios can be measured simultaneously with the cross section.
At the TEVATRON the experiments measure the ratio 
$R=\frac{\sigma(\pp \rightarrow \rm{W} \rightarrow \ell \nu)}
{\sigma(\pp \rightarrow \rm{Z} \rightarrow \ell^+ \ell^-)}$.
The ratio of production cross sections 
$\frac{\sigma(\pp \rightarrow \rm{W})}{\sigma(\pp \rightarrow \rm{Z})
  }$ is then imposed from theory and $\mathrm{Br(Z \rightarrow
  \ell\ell)}$ is taken from LEP, so that $\BrWl$ can be obtained.
The partial width $\Gamma ( \rm{W} \rightarrow \ell \nu)$ is well known from
theory, so that the branching ratio $\BrWl$ measurements can be converted
into $\GW$ measurements. All direct and indirect $\GW$ measurements
are compared with the SM prediction in fig. \ref{fig:gw}. The indirect
measurement from the TEVATRON is slightly high $(1.7 \sigma)$ \cite{ref:gwit}. 
However
the precision of the direct measurements is currently much too bad to
confirm or disprove the disagreement. The LEP number \cite{ref:gwil}
is in between the
SM and the TEVATRON result. However, $\BrWl$ directly is not sensitive
to invisible W decays which should show up in a too small W-pair 
or a too large single W cross section.

\subsection{Measurement of gauge boson couplings}
The most sensitive channel for the measurement of triple gauge boson
couplings is W-pair production at LEP which contains, apart from a
$\nu_e$ t-channel exchange graph also graphs with s-channel $\gamma$
or Z-exchange. However in this channel, contributions from anomalous
ZWW and $\gamma$WW graphs cannot be separated. At the TEVATRON the
most sensitive channel is W$\gamma$ pair production, which is
sensitive to $\gamma$WW couplings only. Some information is also
obtained from WZ and WW pairs. Also at LEP the $\gamma$WW couplings
can be separated using single W production, which is dominated by 
$\gamma$W fusion and single $\gamma$ production.
The experiments usually parametrise the triple gauge couplings with the
five parameters $g_1^Z,\,\kappa_\gamma,\,\kappa_Z,\,
\lambda_\gamma,\,\lambda_Z$ defined in \cite{ref:tgcdef}. The $\gamma$WW and
ZWW couplings are then related by 
$\Delta \kappa_\gamma = 
- \frac{\cos^2 \theta_W}{\sin^2 \theta_W}\left( \Delta \kappa_Z - \Delta g_1^Z 
\right) $
and $\lambda_\gamma =  \lambda_Z $. Table \ref{tab:tgc} summarises the 
single parameter fits for the different 
channels \cite{ref:tgcwwl,ref:tgcwl,ref:tgct}.
Again, no deviations from the predictions are observed. The LEP experiments 
have also performed two parameter fits for any pair of coupling-parameters.
Some of the parameters are significantly correlated, and also in these fits
all results agree with the Standard Model.

\begin{table}[htb]
\begin{center}
\begin{tabular}[c]{|l|c|c|c|}
\hline
 & $\Delta g_1^Z$& $\Delta \kappa_\gamma$& $\lambda_\gamma$ \\
\hline
LEP all & $\pd 0.04 \pm 0.08$ & $-0.01 \pm 0.03$ & $-0.04 \pm 0.04$ \\
LEP single W & & $-0.09 \pm 0.12$ & \\
D0 & & $-0.08 \pm 0.34$ & $\pd 0.00 \pm 0.09$ \\
\hline
\end{tabular}
\end{center}
\caption[bla]{
1-parameter fit results for anomalous triple gauge couplings
}
\label{tab:tgc}
\end{table}

\section{2-fermion physics at high energies}

\begin{figure}[htb]
\begin{center}
\includegraphics[height=9.cm,bb=0 15 567 550]{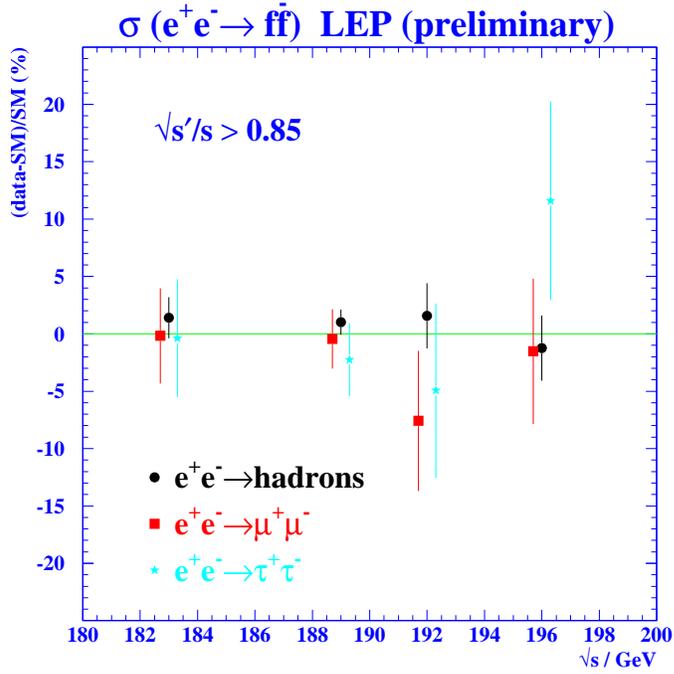}
\end{center}
\caption{
Comparison of the 2-fermion production cross section at LEP2 with the SM 
prediction.
}
\label{fig:sff}
\end{figure}

\begin{figure}[htb]
\begin{center}
\includegraphics[height=8.cm,bb=0 50 567 530]{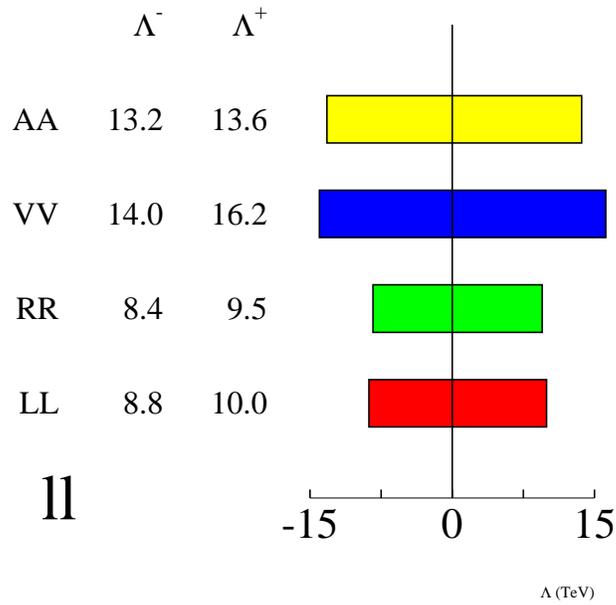}
\end{center}
\caption{
Contact interaction limits (95\%c.l.) at LEP2.
}
\label{fig:contact}
\end{figure}

\begin{figure}[htb]
\begin{center}
\includegraphics[height=9.cm,bb=0 7 567 345]{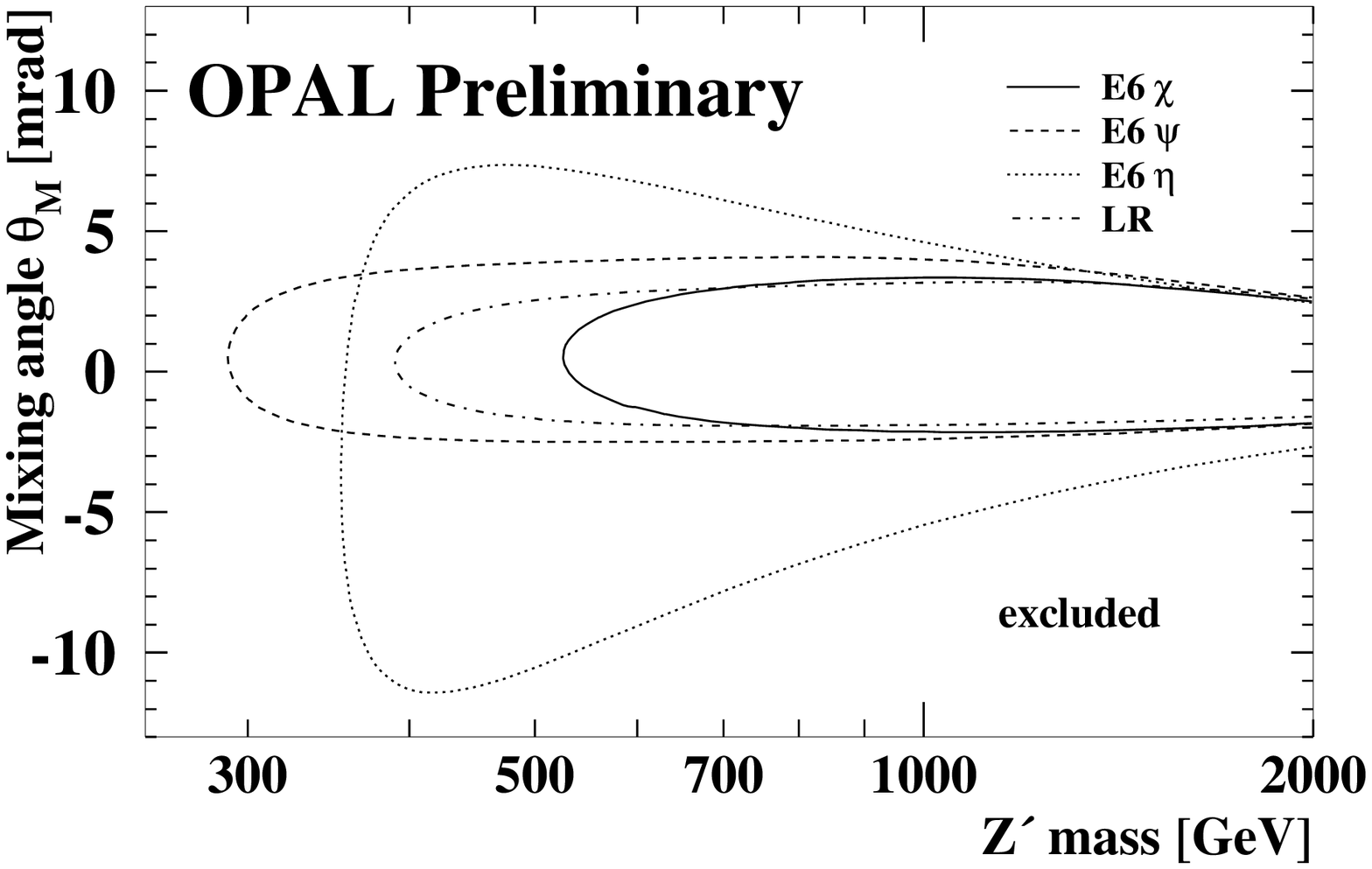}
\end{center}
\caption{
Z' limits (95\%c.l.) for various models from OPAL.
}
\label{fig:zprime}
\end{figure}

2-fermion production at high energies, away from the Z-resonance peak, can be 
used to test a variety of new physics effects. At LEP one has access
to cross sections for lepton- and quark-pair production. For quarks
the total hadronic cross section and the fraction of b-
and c-quarks within the hadronic sample can be measured.
In addition one can measure the forward-backward asymmetry for
leptons, b- and c-quarks.
At energies above the Z, in a large part of the events a high  energy
photon is radiated from the initial state, so that the effective centre
of mass energy of the $\ee$-interaction $(\sqrt{s'})$ is close to the Z-mass.
However, it is no problem to reject these events with high efficiency
using the energy and acolinearity of the measured particles or jets.
As an example figure \ref{fig:sff} shows the comparison of the cross
sections measured at LEP with the Standard Model prediction.
At the TEVATRON a similar process is available with Drell-Yan production of 
lepton pairs, however no new results have been presented in the recent past.
All 2-fermion results agree well with the SM prediction\cite{ref:twofl}.
So they can be used to set limits on a variety of new physics processes.
Figure \ref{fig:contact} shows as an example the limits set on leptonic
contact interaction for different helicity structures \cite{twofint}. 
As another example 
figure \ref{fig:zprime} shows the limits on a possible Z' in different models
obtained by OPAL \cite{ref:zpopal}.
The tight limits on Z-Z' mixing come from the LEP1 precision data while the 
mass limit for zero mixing is from LEP2. For no mixing the TEVATRON sets
limits of comparable size from Drell-Yan production.

\section{Conclusions}
The analysis of the Z-precision data from LEP and SLD is basically finished.
Despite the enormous precision of the data no deviation from the Standard
Model has been found. The data favour a light Higgs boson which is equally
predicted in supersymmetric theories and in the Standard Model if it is
required to be valid up to the Planck scale.
At higher energies at LEP and at the TEVATRON the W-mass has been measured
with comparable precision to the Z-data, leading to the same conclusions.
In addition a variety of other observables have been measured. Since all
agree with the prediction of the electroweak theory stringent limits on new
physics can be set.
\clearpage
\section*{Acknowledgements}
It is a pleasure to thank the organisers of the workshop for the nice and 
friendly
atmosphere at the conference and the conveners and participants of the 
electroweak working group for the useful discussions during the meeting.
Especially I would like to thank Mark Lancaster and Niels Kjaer for their
advice preparing the talk and for reading the manuscript.



\begin{thebibliography}{99}
\bibitem{ref:ew98}
The LEP collaborations, {\it A Combination of Preliminary Electroweak 
Measurements and Constraints on the Standard Model}, CERN-EP/99-15.
%
\bibitem{ref:lepone}
G.~Quast, {\it EW review at LEP1}, 
talk given at EPS 99 Tampere, to appear in the proceedings.
%
\bibitem{ref:zfitter} 
D.~Bardin et al., Z.~Phys. {\bf C44} (1989) 493;
Comp.~Phys.~Comm. {\bf 59} (1990) 303;
Nucl.~Phys. {\bf B351}(1991) 1;
Phys.~Lett. {\bf B255} (1991) 290 and CERN-TH 6443/92 (May 1992).
%
\bibitem{ref:alr}
J.~Brau, {\it Electroweak Precision Measurements with Leptons},
talk given at EPS 99 Tampere, to appear in the proceedings.
%
\bibitem{ref:abac}
S.~Fahey, {\it Measurements of Quark Coupling Asymmetries at the $Z^0$},
talk given at EPS 99 Tampere, to appear in the proceedings.
%
\bibitem{ref:zoltan}
Z.~Kunszt, these proceedings
%
\bibitem{ref:mwh}
R.~Chierici, {\it W mass from fully leptonic and mixed decays at LEP},
talk given at EPS 99 Tampere, to appear in the proceedings.
%
\bibitem{ref:mwl}
L.~Mir, {\it W mass from fully hadronic decays at LEP},
talk given at EPS 99 Tampere, to appear in the proceedings.
%
\bibitem{ref:fsi}
J.~Schiek, {\it New Results on Fragmentation Effects Associated with 
W Production},
talk given at EPS 99 Tampere, to appear in the proceedings.
%
\bibitem{ref:mwt}
B.~Carithers, {\it Measurement of the W Mass at the Tevatron},
talk given at EPS 99 Tampere, to appear in the proceedings.
%
\bibitem{ref:gwl}
D.~Charlton, {\it Precision EW at LEPII},
talk given at LP 99 Stanford, to appear in the proceedings.
%
%
\bibitem{ref:gwit}
CDF Electroweak Group, 
{\it W,Z Cross Sections + W Width Measurement}\\
\verb+http://www-cdf.fnal.gov/physics/ewk/xsec_width_new.html+
%
\bibitem{ref:gwil}
A~.Barczyk, {\it W properties at LEP},
talk given at EPS 99 Tampere, to appear in the proceedings.
%
\bibitem{ref:tgcdef}
K.~Gaemers, G.~Gounaris, Zeit. Phys. {\bf C1} (1979), 259;\\
K.~Hagiwara et al., Nucl. Phys. {\bf B282} (1987), 253.
%
\bibitem{ref:tgcwwl}
A.~Macchiolo, {\it Anomalous couplings from $\ee \rightarrow W^+W^-$},
talk given at EPS 99 Tampere, to appear in the proceedings.
%
\bibitem{ref:tgcwl}
O.~Iouchtchenko, {\it Other anomalous WWV couplings 
(mainly single W production)},
talk given at EPS 99 Tampere, to appear in the proceedings.
%
\bibitem{ref:tgct}
J.~Ellison, {\it W and Z properties at the Tevatron},
talk given at EPS 99 Tampere, to appear in the proceedings.
%
\bibitem{ref:twofl}
M.~Minard, {\it $\ee \rightarrow \ff$ at LEP2},
talk given at EPS 99 Tampere, to appear in the proceedings.
%
\bibitem{twofint}
H.~Rick, {\it Tests of the validity of the SM through $\ee\rightarrow\ff\,,
\ee\rightarrow \gamma s$},
talk given at EPS 99 Tampere, to appear in the proceedings.
%
\bibitem{ref:zpopal}
OPAL Collaboration,
{\it
Limits on a Z' Boson from $\ee$ to fermion pair Cross-sections and Asymmetries
}
OPAL PN-372,
contributed paper to EPS 99 Tampere {\bf HEP'99 19 }.
%
\end{thebibliography}
\end{document}